\documentclass[article,twocolumn,preprintnumbers,nofootinbib]{revtex4-1}
\usepackage{enumerate}
\usepackage{mathrsfs}
\usepackage{amsmath}
\usepackage{amssymb}
\usepackage{natbib}
\usepackage{graphicx}
\usepackage{cancel}
\usepackage[normalem]{ulem}
\usepackage[pdftex,bookmarks,linktocpage,pdfpagelabels,plainpages=false,hyperfigures,linkcolor=blue,citecolor=blue]{hyperref} 
\hypersetup{colorlinks=true}
\usepackage{stackrel}
\usepackage{pgfplots}
\usetikzlibrary{positioning}
\usetikzlibrary{snakes}
\usepackage{tikz-feynman}
\usepackage{dcolumn}
\usepackage{threeparttable}

\newcommand\be{\begin{equation}}
\newcommand\ee{\end{equation}}
\newcommand\bea{\begin{eqnarray}}
\newcommand\eea{\end{eqnarray}}

\begin{document}
\bibliographystyle{apsrev4-1}


\title{Pseudo-Dirac Sterile Neutrino Dark Matter}

\author{Wei Chao$^1$}
\email{chaowei@bnu.edu.cn}
\author{Siyu Jiang$^{1}$}
\email{jiangsy@mail.bnu.edu.cn}
\author{Zhu-Yao Wang$^{2}$}
\email{wang.zhu@northeastern.edu}
\author{Yu-feng Zhou$^{3,4,5}$}
\email{zhouyf@itp.ac.cn}
\affiliation{$^1$Center for Advanced Quantum Studies, Department of Physics, Beijing Normal University, Beijing, 100875, China\\
$^2$Department of Physics, Northeastern University, Boston, MA 02115-5000, USA \\
$^3$CAS key laboratory of theoretical Physics, Institute of Theoretical Physics, Chinese Academy of Sciences, Beijing 100190, China; \\
  School of  Physics, University of Chinese Academy of Sciences, Beijing 100049, China;\\
  $^4$School of Fundamental Physics and Mathematical Sciences, Hangzhou Institute for Advanced Study, UCAS, Hangzhou 310024, China;\\
  $^5$International Centre for Theoretical Physics Asia-Pacific, Beijing/Hangzhou, China}
\vspace{3cm}

\begin{abstract}
Sterile neutrino is a promising dark matter (DM) candidate. However the parameter space of  this scenario has almost been ruled out by the X-ray observation results whenever the sterile neutrino is solely produced by the Dodelson-Widrow (DW) mechanism in the early Universe. In this letter we propose an extension to the minimal sterile neutrino DM model by introducing the pseudo-Dirac sterile neutrino, which implies the existence of two nearly degenerate Majorana states $\hat N_{1,2}^{}$, and a permutation symmetry. The heavy state $\hat N_1$ is produced via the DW mechanism, and the light state $\hat N_2$, which serves as the DM, is produced from the decay of $\hat N_1$ in the early Universe. The X-ray constraint is avoided by the permutation symmetry, which forbidden the two-body decay of the DM into active neutrinos and photon.  A promising signal of this scenario is the effective number of neutrino species, which will be precisely measured in future experiments, such as CMB stage IV. We further study the impact of this model on  the cosmological parameters.  The Markov Chain Monte Carlo analysis for the Planck + BAO+R19 data gives $H_0=69.2_{-0.59}^{+0.58}$, which may relieve the Hubble tension.

\end{abstract}

\maketitle
\section{\bf Introduction} ~The solar, atmosphere, accelerator and reactor neutrino oscillation experiments have confirmed that neutrinos are massive and lepton flavors are mixed~\cite{Xing:2020ijf}, which provide solid evidences of new physics beyond the standard model (SM).  Another evidence of new physics beyond the SM comes from astrophysical observations, which show that about $26.8\%$ of the Universe is made by dark matter (DM)~\cite{ParticleDataGroup:2016lqr}.   There is no cold DM candidate in minimal SM. Various models with DM mass varying from $10^{-22}$ eV up to $10^{\rm 55}$ GeV have been proposed in past few years (for reviews, see \cite{Lin:2019uvt} and references cited therein). However, the non-observation of any signal from the DM direct and indirect experiments has shifted people's interest of DM mass from the electroweak scale to the Sub-GeV scale.  

The keV scale sterile neutrino~\cite{Drewes:2016upu}, proposed to fill the ``desert" spanning six orders of magnitude between ${\cal O}(0.1)$ eV and ${\cal O}(0.1)$ MeV in the lepton mass spectrum, is a well-motivated DM candidate~\cite{Boyarsky:2018tvu}.  Generally, sterile neutrinos can be produced by the neutrino oscillations in the early Universe via a tiny active-sterile neutrino mixing angle and can make up $100\%$ of the DM, which is called Dodelson-Widrow (DW) mechanism~\cite{Dodelson:1993je}. However, such a minimal mechanism is  conflict with the results searching for DM decaying into  monochromatic X-ray~\cite{Abazajian:2017tcc}. There are some  models \cite{Shi:1998km,Nemevsek:2012cd,Shaposhnikov:2006xi,Laine:2008pg,Abazajian:2014gza,Kusenko:2006rh,Bezrukov:2009th,Merle:2013wta} appeared to avoid this constraint. A recent study shows that the X-ray constraint can be avoided by introducing the neutrino self-interactions~\cite{DeGouvea:2019wpf}, $\lambda_\phi \phi  \nu_\alpha \nu_\alpha + {\rm H.c.}$, where $\phi$ is a complex scalar singlet.  However, it has been pointed out in Ref.~\cite{Blinov:2019gcj} that it is difficult to construct such an interaction without violating the electroweak symmetry as the active neutrino is the neutral component of an electroweak doublet and  the Yukawa coupling $\lambda_\phi$ should be proportional to the active neutrino masses~\cite{Kelly:2020aks}.

In this paper we propose a new sterile neutrino DM model, which extends the SM with  a pseudo-Dirac sterile neutrino and a permutation symmetry.   The X-ray constraint may be avoided in this model by setting the sterile neutrino produced from the DW mechanism as an intermediate DM state, which will eventually decay into the second sterile neutrino DM and a dark radiation in the early Universe.   We study the thermodynamics of this model in the  early Universe,  and calculate the effective number of neutrino species, $\Delta N_{\rm eff}^{} $, which is highly correlated with the DM phenomena as the existence of the intermediate sterile neutrino state will slightly modify the evolution dynamics of the standard cosmology.  We point out that the $\Delta N_{\rm eff}^{} $ predicted by this model can be tested by the CMB stage IV, which is expected to reach $\Delta N_{\rm eff}^{} \sim 0.03$. 

We  further investigate impact of this model on the Hubble tension problem, a discrepancy between the CMB measurement~\cite{Planck:2018vyg} of the Hubble constant and the direct measurement in the local Universe from supernovae type Ia~\cite{Riess:2018byc}.  Our result shows that this tension can be relieved in this model due to the decay of the sterile neutrino, which transfers the energy into the radiation and speeds up the expansion of the Universe resulting a larger Hubble constant than the one implied by the $\Lambda$CDM. The Markov Chain Monte Carlo analysis for the Planck +BAO data  and Planck + BAO+R19 data  shows that $H_0=68.31_{-0.61}^{+0.54}$ and $H_0=69.2_{-0.59}^{+0.58}$, respectively. 

The remaining of the paper is organized as follows: In section II we present the model. Section III is devoted to the study the thermodynamics of the early Universe. We discuss the Hubble tension in section IV. The last part is concluding remarks.

\section{Pseudo-Dirac Sterile Neutrino}
In contrast to conventional sterile neutrino model, we assume that sterile neutrino is a pseudo-Dirac particle and its Yukawa interactions satisfy the permutation symmetry, $N_L^{} \leftrightarrow N_R^C$, where $N_L^{}$ and $N_R^{}$ represent the left-handed and right-handed components of the sterile neutrino. This symmetry is similar to the $\mu\leftrightarrow \tau$ symmetry in the neutrino physics~\cite{Xing:2015fdg}.  The relevant Lagrangian can be written as
\begin{eqnarray}
-{\cal L}& = & \hat m \left(\overline{\nu_L^{} } N_R^{}  + \overline{\nu_L} N_L^C \right)+ {1\over 2 }\mu \left( \overline{N_R^C} N_R^{} + \overline{N_L^{} } N_L^C \right) \nonumber \\ &&+ m \overline{N_L^{}} N_R^{} + {\rm h.c.} \label{massterm}
\end{eqnarray}
where $\nu_L^{}$ is left-handed active neutrino,  $\hat m$ is the Dirac mass arising from Yukawa interaction, $ m$ and $\mu$ are the Dirac mass and Majorana mass of sterile neutrinos, respectively.  In the basis $(\nu_L,~ N_L,~ N_R^{} )$, the neutrino mass matrix can be written as
\begin{eqnarray}
M_\nu^{} = \left( \begin{matrix}  \times & \hat m & \hat m \cr  \hat m & \mu & m \cr \hat m  & m & \mu \end{matrix} \right) \; ,
\end{eqnarray}
which can be diagonalized by the $(n_\nu+2)\times(n_\nu+2)$ unitary transformation with $n_\nu$ the generation of active neutrinos, resulting in the mixing between active neutrinos and sterile neutrinos.  The pseudo-Dirac sterile neutrino is then decoupled into two Majorana eigenstates $\hat N_1^{} $ and $\hat N_2^{}$ with a slight mass splitting.  Due to the permutation symmetry,  one of the mass eigenstates, namely $\hat N_2$ with the mass $m_{N_2}^{} =\mu-m$,  which comes from the unitary rotation in the $N_L-N_R^C$ plane with maximal mixing angle $45^\circ$,  does not mix with the active neutrinos.    

The mass of heavier sterile neutrino comes from the diagonalization of the $(n_\nu +1)\times (n_\nu+1)$ sub-mass matrix, which can be roughly written as $m_{N_1}=\mu+m$. In this paper, we assume only one generation left-handed neutrino couples to the sterile neutrino for simplification, but our conclusion  does not change when extended to the three generation case.  In this case the mixing angle  in the $2\times 2$ unitary transformation can easily be calculated as
\begin{eqnarray}
\sin \theta \approx  { \sqrt{2} \hat m \over \mu+m }
\end{eqnarray}
Apparently this mixing angle may lead to the unitarity violation in the PMNS matrix \cite{Maki:1962mu,Pontecorvo:1967fh}, and is thus constrained by the neutrino oscillation data and the meson decay results~\cite{Antusch:2006vwa}.  Given the mixing angle, the neutrino flavor eigenstate can be written in terms of the mass eigenstates
\begin{eqnarray}
\nu_L^{} = \cos\theta \hat \nu_L^{} - \sin \theta\hat N_{1L}^{}  \label{freezein}
\end{eqnarray}
with  $\hat \nu_{}^{} $ and $\hat N_1^{} $  the mass eigenstates of active and sterile neutrinos, respectively.  

Now we consider the interaction between sterile neutrinos in the mass eigenstates.  Due to the permutation symmetry in Yukawa sector,  interactions in terms of $\overline{N_L^{} } \phi N_R^{}  $ and $ \overline{N_L^{} } \Phi N_L^C + \overline{N_R^C } \Phi N_R^{} $, where $\phi$ and $\Phi$ are new scalar singlets,  lead to null interaction between $\hat N_1$ and $\hat N_2$.  An exception is the gauge interaction from a local $U(1)_S$ gauge symmetry,   $g \left( \overline{ N_L^{} } \gamma^\mu N_L^{} + \overline{N_R^{} } \gamma^\mu N_R^{} \right) A_\mu^{\prime} $, where $g$ is the gauge coupling and $A_\mu^{\prime} $ is the new U(1) gauge field.  In the mass eigenstates, one has 
\begin{eqnarray}
{\cal L}_A = -g \sin\theta  \overline{\hat \nu_{}^{} } \gamma^\mu A_\mu^{\prime} P_L^{} \hat N_2^{} - g\cos\theta \overline{\hat N_1^{} }  \gamma^\mu A_\mu^{\prime}  P_L^{} \hat N_2^{}  \; , \label{master}
\end{eqnarray}
which shows  that $\hat N_2$ mainly couple to $\hat N_1$.   In this case, the Majorana mass term in Eq.(\ref{massterm}) comes from the non-zero vacuum expectation value of the scalar singlet that spontaneously breaks the $U(1)_S$ symmetry, while the Dirac mass terms between active and sterile neutrinos come from dimension-5 effective operators~\cite{Chao:2010mp,Cai:2014hka}.

\section{Cosmology}
Due to the existence of pseudo-Dirac neutrino and dark radiation,  the evolution dynamics of the Universe is slightly modified resulting in interesting phenomena. In this section we will study impacts of this model on the DM, effective number of neutrino species and the Hubble tension problems in turn.
\subsection{DM}
As mentioned above, $\hat N_1$ may be produced in the early Universe via neutrino oscillations, it then can decay into $\hat  N_2 $ and $A^\prime$ whenever kinematically allowed.  Here we take $\hat N_2$ as the DM candidate and $A^\prime $ as the dark photon which is super-light due to an extremely small  gauge coupling, similar to that in fuzzy DM model~\cite{Hu:2000ke}.  For the mass of $\hat N_2$, there is a lower limit $m_{\rm DM}> 2$ keV arising from phase space density derived from dwarf galaxies~\cite{Abazajian:2017tcc}.  The Boltzmann equation that describes the evolution of the $\hat N_1$ is 
\begin{eqnarray}
{d  f_{N1}^{}  \over dz } &=& {\Gamma(E,z) \sin^22\theta_{\rm eff}^{} \over 4 Hz}  f_a(E, z) \Theta(E-m_{N1}^{}) \nonumber \\ 
&-& {f_{N1}^{}  \over   Hz} {m_{N1} \over E} \Gamma(\hat N_1 \to \hat N_2 + A^\prime) \label{b-DM}
\end{eqnarray}
where the first term on the right-handed side is from the neutrino oscillation\cite{Abazajian:2001nj} and the second term is from the decay of $\hat N_1$, $f_{N1}^{}$ and $f_\alpha$  are the phase space distribution function of  the sterile neutrino $\hat N_1$ and active neutrino respectively, $\Theta(x)$ is the Heavy side function, $\theta_{\rm eff}^{} $ is the effective mixing angle~\cite{Abazajian:2005gj,Hansen:2017rxr}. The distribution function  of active neutrinos is characterized by the temperature $T$ and the chemical potential $\mu_\nu$,
\begin{eqnarray}
f_a = {1\over 1+ \exp\left({E-\mu_\nu\over T}\right)}\; .
\end{eqnarray}
Interactions contributing to $\Gamma(E, z)$ include $\nu_a \nu_\beta \leftrightarrow \nu_a \nu_\beta$, $\nu_a \ell^\pm \leftrightarrow \nu_a \ell^\pm$, $\nu_a q \leftrightarrow \nu_a q$ and $\nu_a \nu_a \leftrightarrow \ell^\pm \ell^\mp$. The total interaction rate due to neutrino self interactions and interactions with electron positron pairs is~\cite{Abazajian:2001nj}
\begin{eqnarray}
\Gamma(E, z) \approx \left\{ \begin{matrix} 1.27 G_F^2 E T^4 & a =e \cr 0.92 G_F^2 E T^4 & a=\mu,\tau \end{matrix}\right.
\end{eqnarray}
where $G_F$ is the Fermi constant. At higher temperatures, $\mu^\pm$, $\tau^\pm$ and quarks contribute to the neutrino interaction rate, which will be included in our calculation.  

\begin{figure}[t]
  \centering
  \includegraphics[width=0.48\textwidth]{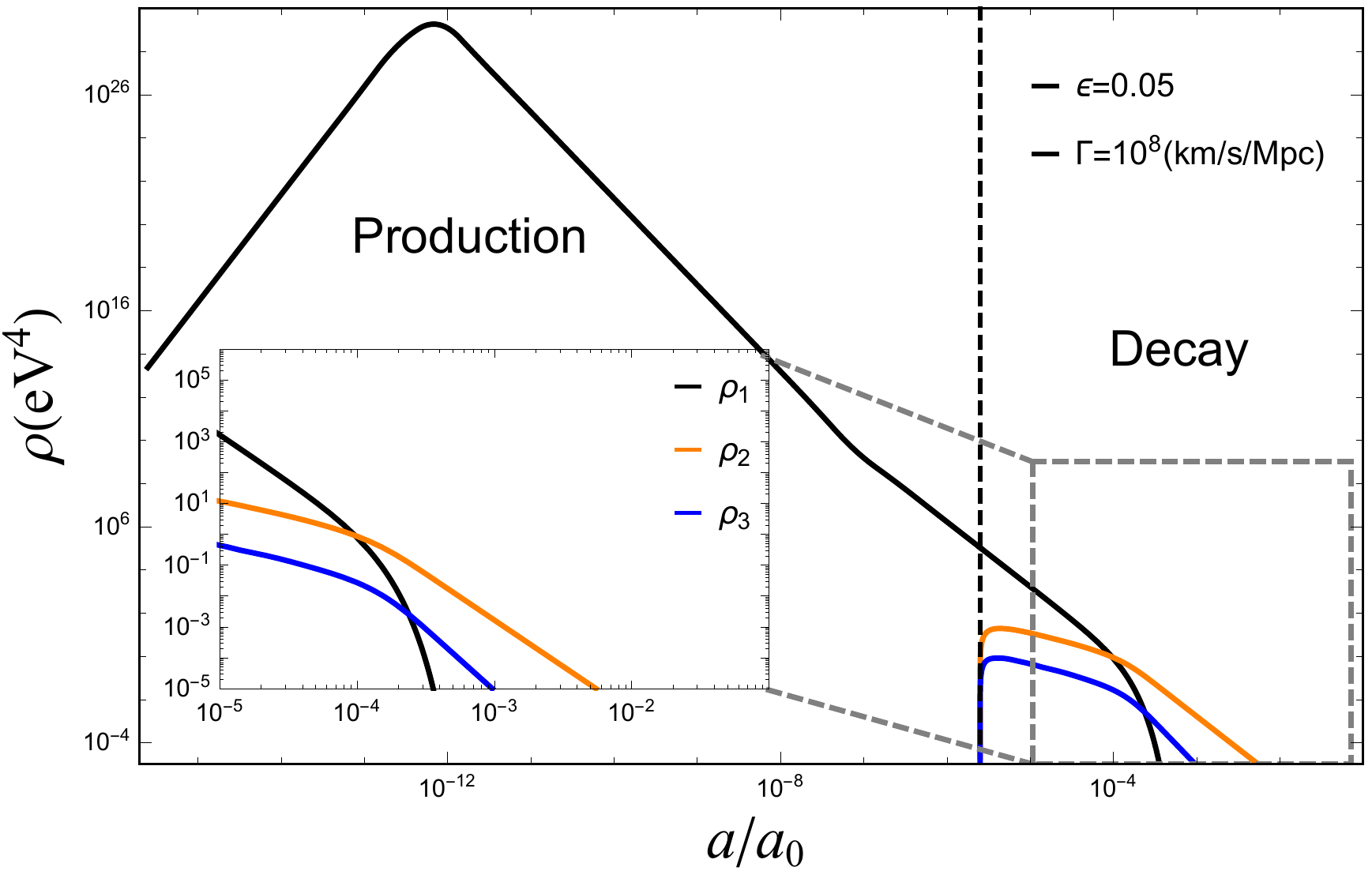}
\caption{ The energy densities of the heavy sterile neutrino $\rho_1$, light sterile neutrino DM $\rho_2$ and the dark radiation $\rho_3$ as the function of the scale factor $a/a_0$ where $a_0$ is the present scale factor.}
\label{e_densities}
\end{figure}

The DM $\hat N_2$ is produced via the freeze-in mechanism~\cite{Hall:2009bx}  with the Boltzmann equation
\begin{eqnarray}
{d  f_{N2}^{}  \over dz } &=&{f_{N1}^{}  \over   Hz} {m_{N1} \over E} \Gamma_1^{}(\hat N_1 \to \hat N_2 + A^\prime) \nonumber \\ 
&-& {f_{N2}^{}  \over   Hz} {m_{N2} \over E} \Gamma_2^{} (\hat N_2 \to \hat \nu + A^\prime) \label{true-DM}
\end{eqnarray}
Combing the Eq.(\ref{b-DM}) with Eq.~(\ref{true-DM}), one can solve the number density of DM numerically to get the final relic abundance. The decay rate of sterile neutrinos can be written as
\begin{eqnarray}
\Gamma_1^{} (\hat N_1 \to \hat N_2 + A^\prime) &=& {1 \over 4 } \alpha_X^{} m_{N_1} \cos^2 \theta (1-\xi^2 )\\
\Gamma_2^{}(\hat N_2 \to \hat \nu + A^\prime) &=& {1\over 4}  \alpha_{X}^{} m_{N_2}^{}   \sin^2 \theta 
\end{eqnarray}
where we have assumed that masses of $\hat \nu$ and $A^\prime$ are negligible compared to those of sterile neutrinos, $\alpha_X= g^2 /4\pi$ and $\xi\equiv 2\varepsilon=1-m_{N_2}^2/m_{N_1}^2$.  The decay rate of $ N_2$ is suppressed by the factor of $\tan^2 \theta$ compared with the decay rate of $N_1^{} $.  

We show in the Fig.~\ref{e_densities}, energy densities of $\hat N_1$ , $\hat N_2$ and $A^\prime$, i.e. $\rho_1$, $\rho_2$, $\rho_3$,  as the function of scale factor $a/a_0$, where $a_0$ is the present day scale factor, by setting $\sin^{2}\theta=7\times 10^{-11}$, $\hat m_{N_1} =7~{\rm keV}$, $\varepsilon=0.05$ and $\Gamma_{\mathrm N_1}=10^8 ~{\rm km/s/Mpc}$.  One can find that $\hat N_1$ is produced relativistically via neutrino oscillation in the early Universe  at about ${\cal O}(100)~{\rm MeV}$. It subsequently  decays into $\hat N_2$ and $A^\prime$. Given $\rho_{2}^{}$ one can easily derive the relic density of DM today, $\Omega=\rho_2/\rho_C$ where $\rho_C= 1.05\times 10^{-5} h^{-2} ~{\rm GeV/cm^3}$ being the critical density.  Alternatively, the observed value $\Omega h^2 =0.12$ may applied to constrain the parameter space of the model.  Since $\rho_1$ decayed away in the early Universe and $\rho_2$ does not decay into photon, constraint from X-ray observation can be avoided.

\begin{figure}[t]
  \centering
  \includegraphics[width=0.49\textwidth]{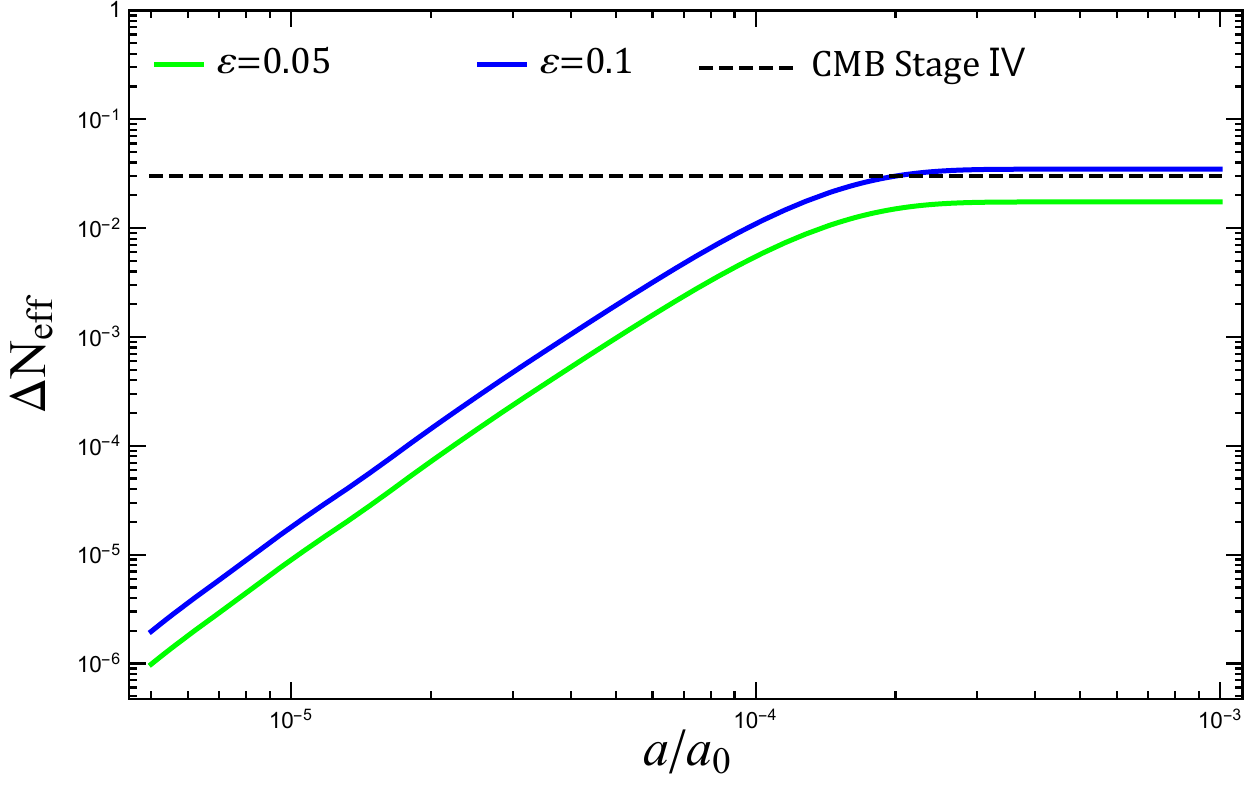}
\caption{ The  effective number of neutrino species as the function of the scale factor $a/a_0$.}
\label{neff}
\end{figure}

\subsection{The effective number of neutrino species}
The existence of  new physics relevant to neutrino could modify the effective number of relativistic neutrino species $N_{\rm eff}$, which is an important cosmological parameter and is stringently constrained by the precision data from CMB observations.   Planck 2018 gives, $N_{\rm eff}=2.99\pm0.17$~\cite{Planck:2018nkj,Planck:2018vyg}, which is consistent with the SM prediction $N_{\rm eff}^{\rm SM} =3.045$~\cite{Mangano:2005cc,Grohs:2015tfy,deSalas:2016ztq,EscuderoAbenza:2020cmq}. Recent studies show that a variety of BSM neutrino physics scenarios, including non-standard neutrino interactions,  may significantly contribute to $N_{\rm eff}^{}$ \cite{Escudero:2019gzq,Kelly:2020aks}. These scenarios will be tightly constrained by the future measurement from experiments such as CMB Stage IV, which are expected to reach a precision of $\Delta N_{\rm eff}^{}  =N_{\rm eff}^{} -N_{\rm eff}^{\rm SM} \sim 0.03$~\cite{CMB-S4:2016ple,Abazajian:2019eic}.

The temperature relation between neutrinos and photons after the neutrino decoupling can be  derived from the entropy conservation arguments, $T_\nu=(4/11)^{1/3} T_\gamma^{} $.  The radiation energy density is then the sum of energy density of photon $\rho_\gamma$, neutrino $\rho_\nu $ and dark radiation $\rho_{\rm DR}^{} $
\begin{eqnarray}
\rho_{\rm R} = \rho_\gamma + \rho_\nu + \rho_{\rm DR} = \left[ 1+ {7\over 8} \left( {4\over 11}\right)^{4/3} N_{\rm eff}\right] \rho_\gamma \; , 
\end{eqnarray}
where $N_{\rm eff}^{} $ is the effective number of neutrino species with the default value  $3$ by definition, corresponding to three generation of active neutrinos.   More precisely, any non-standard energy density can act as $N_{\rm eff}^{}$, which can be written as
\begin{eqnarray}
\Delta N_{\rm eff}^{} = {8\over 7} \left(  {11\over 4}\right)^{4/3} {\Delta \rho \over \rho_\gamma^{} }
\end{eqnarray} 
where $\Delta \rho$ is the deviation of radiation energy density from the SM prediction. In our model, $\Delta \rho $ comes from decay of the heavy sterile neutrino. $\Delta N_{\rm eff}$ is measured both at the big bang nucleosynthesis (BBN)~\cite{Cyburt:2015mya} and CMB epochs.  At the BBN epoch, one has 
$
\Delta N_{\rm eff}^{\rm BBN} \sim 3.046\times {\rho_R (T^{\rm BBN})/ \rho_\nu^{} (T^{\rm BBN})} 
$
where the pre-factor is the SM prediction of $N_{\rm eff}^{}$.  Due to a tiny gauge coupling $g$,  the heavy sterile neutrino decays after the BBN epoch, so that its impact to the  $\Delta N_{\rm eff}^{\rm BBN}$ is negligible and one only needs to concern $\Delta N_{\rm eff}$  at the CMB epoch.

The Boltzmann equations for the energy density of   dark radiation $A^\prime$  and active neutrino  are 
\begin{eqnarray}
\dot{\rho}_{\rm DR}^{} + 4 H\rho_{\rm DR}^{} &=&  \varepsilon \Gamma_1 \rho_{1}^{} + {1\over 2} \Gamma_{2}^{}   \rho_{2}^{} \\
\dot{\rho}_\nu + 4 H \rho_\nu^{} &=&  {1\over 2} \Gamma_{2}^{}   \rho_{2}^{} 
\end{eqnarray}
where $\rho_\nu$, $\rho_{\rm DR}$, $\rho_{1}$ and $\rho_{2}$ are energy density of active neutrino,  $A^\prime$,  $\hat N_1^{}$ and $\hat N_2^{}$ respectively, $H$ is the Hubble parameter, $\Gamma_{1,2}$ is the decay rates given in Eqs.(10) and (11),  $\varepsilon$ being the fraction of energy transferred into dark radiation, $\varepsilon \approx 1-m_{N_2}/m_{N_1}$.

Fig.~\ref{neff} shows the $\Delta N_{\rm eff}$ as the function of the scale factor $a/a_0$. The green and blue lines correspond to $\varepsilon=0.05$ and $0.1$ respectively.  The dashed horizontal line  represent the reach of future CMB stage IV experiment. It shows that this model can be indirectly tested by the precision measurement of $N_{\rm eff}^{}$.

\begin{figure}[t]
  \centering
  \includegraphics[width=0.49\textwidth]{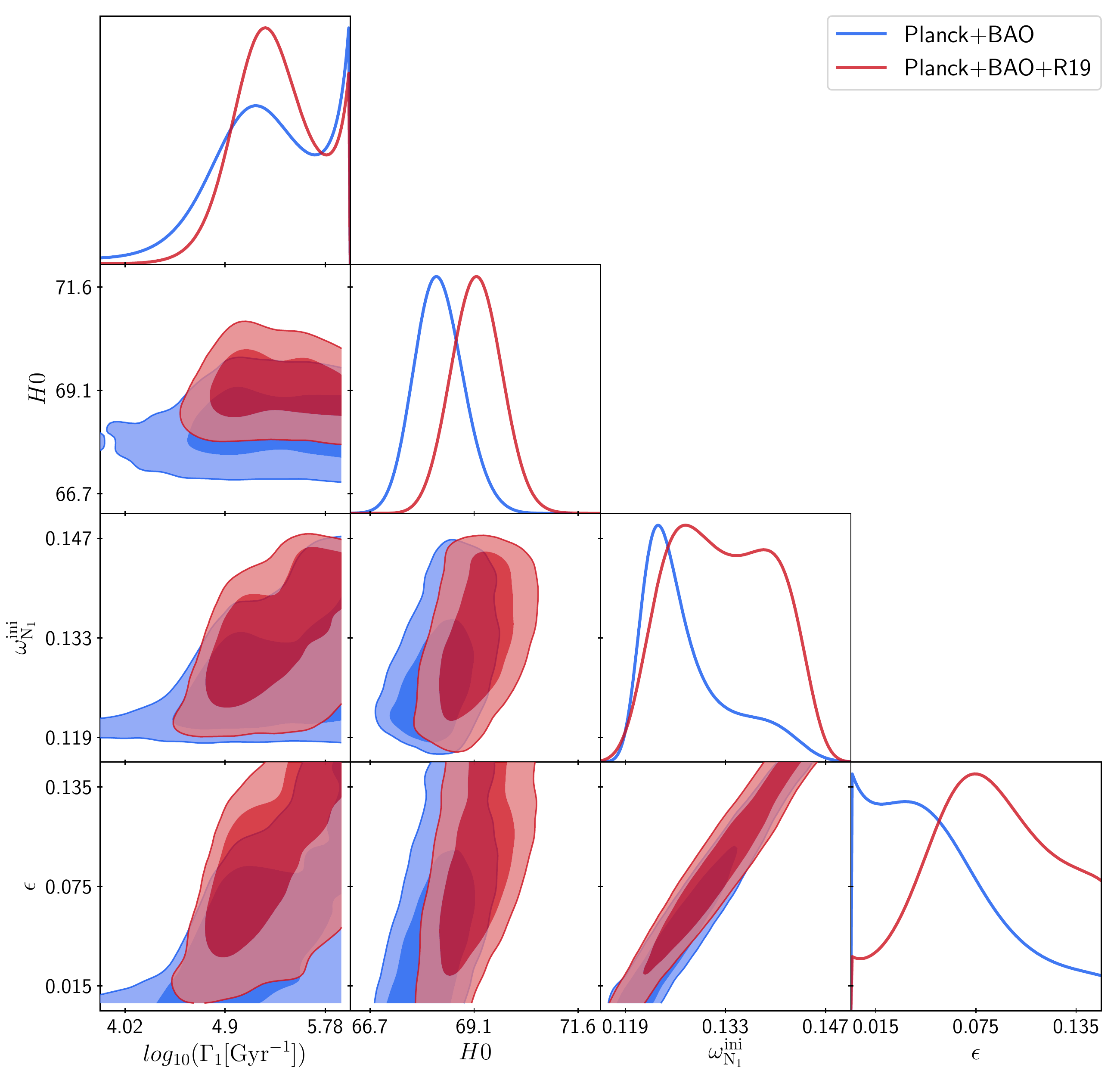}
\caption{ Marginal constraints on the fitted cosmological parameters  for Planck + BAO+R19 (red) and Planck +BAO (blue).  Likelihood contours show the 1$\sigma$ and $2\sigma$ confidence levels.}
\label{hubble}
\end{figure}

\subsection{The Hubble tension}

The Hubble constant, which measures the expansion rate of the Universe, is one of most important cosmological observables. The Hubble Space Telescope (HST) gives a precise estimation of $H_0$, $H_0=(73.42\pm 1.74) $ ${\rm km~s^{-1}~Mpc^{-1}}$~\cite{Riess:2019cxk,Riess:2021jrx}. Alternatively, the Planck satellite data has measured the temperature and polarization anisotropies assuming  a $\Lambda$CDM model, resulting in  $H_0=(67.37\pm0.54)$ ${\rm km~s^{-1}~Mpc^{-1}}$ \cite{Planck:2018vyg}. These two independent measurements of $H_0$ are in tension with each other. Attempts to relieve the tension include modifying the dark energy equation of state\cite{Poulin:2018cxd,Karwal:2016vyq,DiValentino:2017zyq} or the DM model\cite{Ko:2017uyb,DEramo:2018vss,Kumar:2019wfs,Berezhiani:2015yta,Vattis:2019efj}. For a recent review see, Ref.~\cite{DiValentino:2021izs,Shah:2021onj}. In this section, we will address the Hubble tension problem by using the decay of $\hat N_1$.  

In our model, the energy conservation for DM and dark radiation fluids~\cite{Kang:1993xz,Audren:2014bca}  yields
{\small
\begin{eqnarray}
\rho_{1}^{} (a)&=& \rho_{*}^{}  e^{-\Gamma_1^{} [t-t_*] }  \left( {a_{*}^{} \over a } \right)^3 \\   
\rho_2^{} (a) &\approx & {\rho_*} (1-\varepsilon) \left( 1- e^{-\Gamma_1 (t-t_*)}\right)  \left( {a_* \over a }\right)^3\\
\rho_{\rm DR}^{} (a) &=& {\varepsilon \rho_{*}^{}  a_{*}^3 e^{\Gamma_1^{} t_* } \over a^4} \left[ \int_{a_*}^a e^{-\Gamma_1 t } da - \left. a e^{-\Gamma_1^{}  t} \right |_{a^*}^a\right] 
\end{eqnarray}
}where $a_*$ is the scale factor when $N_1$ start to decay, $\rho_*$ is the energy density of $\hat N_1$ at $a_*$, $t=t(a)$ and $t_*=t(a_*)$. In deriving the energy density of $\hat N_2$, we have neglected the impact of $\Gamma_2$, which is tens of order smaller than $\Gamma_1^{}$.

The dark radiation from the decay of $\hat N_1$ will modify the expansion of the Universe  and thus may relieve the Hubble tension, which is similar to the case of  decaying DM.   On the other hand, the fitted cosmological parameters may put constrain on the parameter space of this model.   We use the Friedmann equation for a flat geometry,
\begin{eqnarray}
H^2 (a) =\left( \dot{a} \over a \right)^2 = {8\pi G \over 3 }\hat \rho (a)
\end{eqnarray}
where $\hat \rho(a) =\rho_{1}^{} (a) +\rho_{2}^{} (a) + \rho_{\rm DR}^{} (a) + \rho_\Lambda +\rho_b  + \rho_\gamma(a)+ \rho_\nu(a)$ with $\rho_\Lambda$ the energy density of dark energy and $\rho_b$  the energy density of baryons, to perform a Markov Chain Monte Carlo (MCMC) analysis.

In order to obtain the CMB constraint on cosmological parameters we use the public code CLASS \cite{Blas:2011rf} and MontePython \cite{Audren:2012wb} to run MCMC using the following set of cosmological parameters \cite{Planck:2018vyg}:
\begin{eqnarray}
	\Theta = \{ \omega_{b },H_0,\ln(10^{10}A_{s }),n_{s },\tau_{reio },\sigma_8 \}
\end{eqnarray}
in addition to our model dependent parameters  $\{\omega^{\rm ini}_{\mathrm N_1},\varepsilon, \Gamma_{1} \}$, where $\omega_{i} (i=b, ~{\rm N_1})$ represents $\Omega_{i}h^{2}$.  We set $ \varepsilon$ and  $\Gamma_{1}$ in the following range: $ \varepsilon \in[0,0.15], ~\log_{10}(\Gamma_{1}/\mathrm{Gyr}^{-1}) \in[4,6] $,
such that $\hat N_1$ decay well before the recombination and $\hat N_2$ can also be viewed as cold DM. Actually, CMB is insensitive to the clustering properties of DM~\cite{Voruz:2013vqa}.
And we further use the following datasets to analyze the model:
\begin{itemize}
\item The 2018 Planck measurements of the CMB \cite{Planck:2019nip} (via TTTEEE Plik  high-l, TT and EE low-l, and lensing likelihoods).
\item Baryonic Acoustic Oscillation data (BAO) from BOSS DR12 \cite{BOSS:2016wmc}.
\item the local measurement of $H_0$ from SH0ES (R19) \cite{Riess:2019cxk}.
\end{itemize}

\begin{table}[t!]
		\centering
	\begin{threeparttable}
		\begin{tabular}{|l|c|c|c|c|} 
			\hline 
			&  $\Lambda$CDM & DNDM & DNDM\\
			&  Planck+BAO        &Planck+BAO & Planck+BAO+R19 \\
			Param  &  mean$\pm\sigma$ & mean$\pm\sigma$ & mean$\pm\sigma$\\ \hline 
			$100~\omega_{b }$  & $2.242_{-0.014}^{+0.013}$ & $2.241_{-0.016}^{+0.016}$ & $2.258_{-0.015}^{+0.015}$ \\ 
			$\log_{10}(\Gamma_{1})$ &- & $\textgreater 5.199$ & \textgreater 5.32  \\ 
			$H_0$ & $67.7_{-0.45}^{+0.45}$ & $68.31_{-0.61}^{+0.54}$ & $69.2_{-0.59}^{+0.58}$ \\ 
			$\ln10^{10}A_{s }$ & $3.049_{-0.015}^{+0.014}$ & $3.056_{-0.016}^{+0.014}$ & $3.065_{-0.017}^{+0.016}$\\ 
			$n_{s }$ & $0.9664_{-0.004}^{+0.0039}$ & $0.973_{-0.0076}^{+0.0053}$ & $0.9805_{-0.0086}^{+0.0064}$ \\ 
			$\tau_{reio }$ & $0.05718_{-0.0076}^{+0.0072}$ & $0.05774_{-0.008}^{+0.0068}$   & $0.06155_{-0.0083}^{+0.0072}$\\ 
			$\omega^{\mathrm{ini}}_{\mathrm{N}_{1}}$ &$0.1194_{-0.00099}^{+0.00098}$ & $0.1278_{-0.0086}^{+0.0032}$  & $0.1329_{-0.0076}^{+0.0071}$ \\ 
			$\varepsilon$ &- & $\textless 0.06889$ & $0.09233_{-0.048}^{+0.04}$ \\ 
			$\sigma_8$ & $0.8104_{-0.0062}^{+0.0063}$ & $0.8189_{-0.01}^{+0.0077}$ & $0.8243_{-0.012}^{+0.0096}$ \\ 
			\hline 
		\end{tabular}
		\caption{Table for cosmological parameters}\label{tab:tablenotes}
	\end{threeparttable}
\end{table}

In fig.\ref{hubble} we  show the 2D-plot of ($\omega^{\rm ini}_{\rm N_1}$, $\varepsilon$, $\log(\Gamma_{1}/\mathrm{Gyr}^{-1})$, $H_0$) by using Planck+BAO and Planck+BAO+R19. We can see that the decay of $\hat N_1$ can relieve the Hubble tension. The Hubble constant is positively related to the parameter $\varepsilon$, which controls the energy density of the dark radiation and thus $\Delta N_{\rm eff}^{} $. Enlarged $N_{\rm eff}^{} $ results in a smaller sound horizon. Thus larger Hubble constant is needed to maintain the consistency with the CMB power spectrum.
We dub our model as Decaying Neutrino DM (DNDM) and  summarize in Tab.~\ref{tab:tablenotes} constraints on various cosmological parameters. The data of $\Lambda$CDM is included for comparing, in which $\omega_{\rm cdm}=\omega_{\rm N_1}$ with  $\hat N_1$ the  stable DM. We can see that $\epsilon < 0.069$ for Planck+BAO. For Planck+BAO+R19, constraint on $\varepsilon$ is weaken because of the large prior on $H_0$.  Just like other works \cite{Abellan:2021bpx,Nygaard:2020sow,Xiao:2019ccl,Pandey:2019plg,Poulin:2016nat}, the decay DM prefers small share of dark radiation. It is natural in our model because of the Pseudo-Dirac property of the sterile neutrino. Alternatively, we can use the dark radiation density to constrain the Pseudo-Dirac property. It should be pointed out that, if the mass of $\hat N_1$ is smaller, i.e. it behaves like warm DM before recombination as in Ref.~\cite{Blinov:2020uvz}, the model will work better because of the sufficiently long free-streaming length of sterile neutrino. However it can't make up all the DM due to other cosmological bounds.

\section{Conclusion}
Sterile neutrino is a typical warm DM candidate at the keV scale. However most of its parameter space has been ruled-out  by the X-ray observations. We have explored one possibility of avoiding the X-ray  constraint by introducing the concept of the  pseudo-Dirac sterile neutrino with permutation symmetry in the Yukawa interaction sector.  Our result shows that the light component  of sterile neutrino, which originates from the decay of the heavy sterile neutrino state, can be successful DM candidate in this scenario.  
In contradict to traditional sterile neutrino DM,   $\hat N_{\hat 1,2}$ mainly decay into dark radiation instead of photon, such that X-ray constraint can be avoided. Importantly, the effective number of neutrino species serves as an indirect detection signal of this model. Impact of  the model on the cosmological parameters are explored. Typically the MCMC analysis shows that the Hubble tension problem can be relieved.

\begin{acknowledgments}
The authors thank to Dr. Andreas Nygaard and Dr. Guillermo Franco Abellan for helpful discussions.  WC is supported by the National Natural Science Foundation of China under grant No. 11775025, No. 12175027, and the Fundamental Research Funds for the Central Universities under grant No. 2017NT17.  ZYF is supported in part by the National Key R\&D Program of China No.~2017YFA0402204, the Key Research Program of the Chinese Academy of Sciences (CAS), Grant NO.~XDPB15, the CAS Project for Young Scientists in Basic Research YSBR-006, and
the National Natural Science Foundation of China (NSFC) No.~11825506,   No.~11821505,
and No.~12047503.
\end{acknowledgments}



\bibliography{Sterile}


\end{document}